\documentclass[oldversion]{aa}
\usepackage{graphicx}
 \usepackage{txfonts}
\begin{document}
   \title{The Plasma Emission Model of RX J1856.5-3754}

   \subtitle{}

   \author{N. Chkheidze
          \inst{1}
          \and
          G. Machabeli\inst{2}
          }
\offprints{Nino Chkheidze }

   \institute{Tbilisi State University, Chavchavadze Avenue 3, 0128,
Tbilisi, Georgia\\
              \email{n.chkheidze@gmail.com}
         \and
             Abastumani Astrophysical Observatory, Al. Kazbegi Avenue 2a,
0160, Tbilisi, Georgia\\
             \email{g.machabeli@astro-ge.org}
            }

   \date{Received / accepted}

 \abstract{A spectral analysis of the nearby isolated
neutron star RX J1856.5-3754 is presented. Applying the kinetic
approach, the distribution functions of emitting electrons are
derived and the entire spectra is fitted. It is found that waves
excited by the cyclotron mechanism, come in the radio domain. We
confirm that the cyclotron instability is quite efficient, since the
estimations show that the time of wave-particle interaction is long
enough for particles to acquire perpendicular momentum and generate
observed radiation. The lack of rotational modulation is discussed
and the pulsar spin period is estimated. }

\maketitle
%

\newpage

\section{Introduction}

The isolated neutron star RX J1856.5-3754 (hereafter RXJ1856) was
discovered by ROSAT as an X-ray source (\cite{walt96}). According to
the observational evidence the emission of RXJ1856 does not show any
significant periodic variations and the featureless X-ray spectrum
can be fit by the Planckian spectrum with a temperature
\(kT_{bb}^{\infty}\simeq63\pm3\)eV (Burwitz et al. 2003). It has
been proposed that the emission of this source has a thermal nature
and every effort is bent to make a model which would well describe
the overall spectra.

First spectral modeling of RXJ1856 has been presented by
\cite{pavlov96}. It was shown that the light element (hydrogen or
helium) nonmagnetic NS atmosphere models can be firmly ruled out,
because they over-predict the optical flux by a large factor. On
the other hand no acceptable fit can be obtained with iron and
standard solar-mixture atmosphere models, because the features
predicted by these models are not detected with high significance.
Doppler smearing of the spectral lines due to fast rotation of a
NS does not wash away completely the strongest spectral features
(\cite{pavlov02}, \cite{braje02}).

The similar problems appear for highly magnetized NS atmosphere
models (\cite{raja97}, \cite{zav02}). So one has to conclude that
the classic NS atmosphere models are unable to explain the observed
X-ray emission of RXJ1856.

Different explanation have been considered in papers: \cite{burw01},
\cite{turolla04}. It has been proposed\footnote{Originally suggested
by G. Pavlov (2000)} that the star has no atmosphere but a condensed
matter surface. The mentioned surface might result in a virtually
featureless Planckian spectrum in the soft X-ray band. But yet
another problem arose from the fact that the parameters derived from
X-rays do not fit the optical spectrum with an intensity 6 times
larger, than that of a X-ray emission. This situation led
\cite{pons02} to introducing the overall spectra by two components.
In this model the soft-component of \(kT_{bb}^{\infty}\simeq20\)eV
represents radiation from relatively cool surface and fit the
optical data, when the hard component of
\(kT_{bb}^{\infty}\simeq55\)eV emitted from \(\sim20\%\) of the NS
surface is being responsible for the X-ray emission.

According to \cite{burw03}, the simplest way to produce a time
constant flux would be to assume a uniform temperature distribution
across the stellar surface. In this model the optical data match the
Planckian spectrum for a blackbody temperature
\(kT_{bb}^{\infty}\simeq63\)eV and a radius
\(R_{bb}^{\infty}\simeq12.3\)km respectively, but the X-ray
emissivity is below that of a blackbody by a substantial factor. To
explain this fact and also fit the overall spectra, \cite{burw03}
supposed that the radiating surface should have a high reflectivity
in the X-ray domain.

So to make these models work, it has been postulated that the star
has a condensed matter surface. The existence of such a medium
requires low temperature and strong magnetic field (\cite{lai97},
\cite{lai01}). The former seems to be fulfilled for both models
(Pons et al. 2002, Burwitz et al. 2003), when the magnetic field of
the star is still unknown. There are no convincing theories of
existence of such a medium, in which it would be possible to confirm
the required non-uniform distribution of the surface temperature
(two-component blackbody model). Also it should be shown that such
surface has a required reflectivity in the X-ray domain, which is
one of the main conditions in the model of \cite{burw03}.  Existing
models, based on an assumption that the emission is of thermal
nature, have a number of problems.

\begin{figure}
   \centering
\includegraphics[width=4cm]{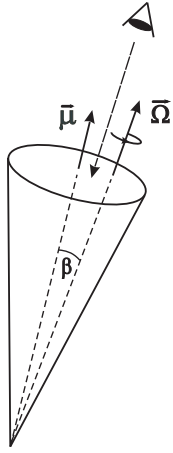}
      \caption{{\bf $\Omega$} is pulsar spin-axis,$\ \mu$ -axis of
magnetic-momentum (\(\beta\ll1^{\circ}\))
              }
         \label{Fig1}
   \end{figure}

We are not about to reject the existing models, but in present paper
we propose our explanation of emission of RXJ1856, based on well
developed models of pulsars. The lack of detected spin-modulation is
easily explained by assumption of a nearly aligned rotator (see
Fig.~\ref{Fig1}), in this case the line of sight of an observer
always lies in the emission cone and the observer receives
continuous radiation. We will show below that the presented model
does not face with a problem of the absence of spectral features in
the observed radiation. The lack of the spectral lines imposes very
stringent constraints on thermal radiation models. However, a recent
paper by Ho et al. (2007) explains the observed featureless spectra
of RXJ1856, based on an assumption that the star has a thin
magnetic, partially ionized atmosphere on the top of the condensed
surface. Though creation of thin hydrogen atmospheres appears to be
one of important problems for this model. To explain the measured
spectra we refer to the pulsar emission model proposed by
\cite{lom79} and by Machabeli \& Usov (1979), where it has been
shown that the  waves excited by the cyclotron instability interact
with particles, leading to the appearance of pitch-angles, which
obviously causes synchrotron radiation.

In this paper, we describe the emission model (\S2), derive X-ray
(\S3) and optical (\S4) spectra, based on the emission model
presented in \S2, confirm the efficiency of the cyclotron mechanism
(\S5) and make conclusions (\S6).

%

\section{Emission model}

%
%

\begin{figure}
   \centering
\includegraphics[width=6cm]{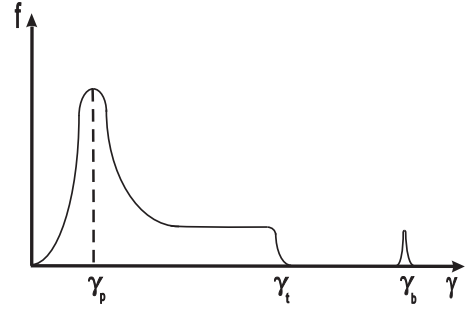}
      \caption{Distribution function of a
one-dimensional plasma in the pulsar magnetosphere. Left corresponds
to secondary particles, right to the primary beam.
              }
         \label{Fig2}
   \end{figure}

\begin{figure}
   \centering
\includegraphics[width=6cm]{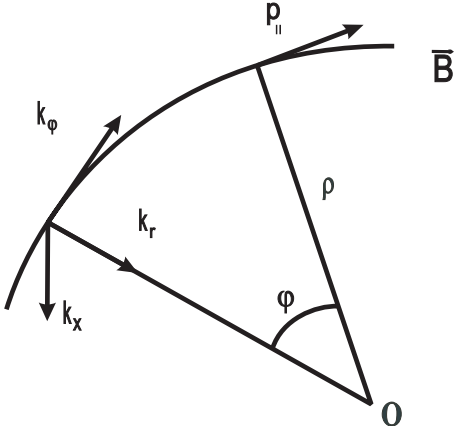}
      \caption{Coordinate system (\(\rho\) is a local radius of curvature)
              }
         \label{Fig3}
   \end{figure}

It is well known that the distribution function of relativistic
particles is one dimensional at the pulsar surface, because any
transverse momenta (\(p_{\perp}\)) of relativistic electrons are
lost during a very short time(\(\leq10^{-20}\)s) via synchrotron
emission in very strong \(B\sim10^{11}\)G magnetic fields. The
distribution function is shown in Fig.~\ref{Fig2} (\cite{arons81}).
For typical pulsars the plasma consists of the following components:
the bulk of plasma with an average Lorentz-factor
\(\gamma_{p}\simeq10\); a tail on the distribution function with
\(\gamma_{t}\simeq10^{4}\) and the primary beam with
\(\gamma_{b}\simeq10^{6}\). Though, plasma with an anisotropic
distribution function becomes unstable that can lead to a wave
excitation in the pulsar magnetosphere. The main mechanism of wave
generation in plasmas of the pulsar magnetosphere is the cyclotron
instability. The process can be conveniently described in
cylindrical coordinates (see Fig.~\ref{Fig3}). The cyclotron
resonance condition can be written in the form (\cite{kaz91}):
\begin{equation}\label{1}
    \omega-k_{\varphi}V_{\varphi}-k_{x}u_{x}+\frac{\omega_{B}}{\gamma_{r}}=0,
\end{equation}
where \(V_{\varphi}\approx
c(1-\frac{u_{x}^{2}}{c^{2}}-\frac{1}{2\gamma_{r}^{2}})\),
\(k_{\varphi}^{2}+k_{\perp}^{2}=k^{2}\),
\(k_{\perp}^{2}=k_{x}^{2}+k_{r}^{2}\),  \(\gamma_{r}\) is the
Lorentz-factor for the resonant particles, \(u_{x}\) is the drift
velocity of the particles due to curvature of the field lines:
\begin{equation}\label{2}
    u_{x}=\frac{cV_{\varphi}\gamma_{r}}{\rho\omega_{B}},
\end{equation}
where \(\rho\) is the radius of curvature of the field lines and
\(\omega_{B}=eB/mc\) is the cyclotron frequency. The resonant
condition for transverse (t) waves with the spectrum:
\begin{equation}\label{3}
    \omega_{t}=kc(1-\delta),
\end{equation}
where
\begin{equation}\label{4}
    \delta=\frac{\omega_{p}^{2}}{4\omega_{B}^{2}\gamma_{p}^{3}},
\end{equation}
takes the form:
\begin{equation}\label{5}
    \frac{1}{2\gamma_{r}}+\frac{\left(k_{\perp}/k_{\varphi}-u_{x}/c\right)^{2}}{2}+\frac{1}{2}\frac{k_{r}^{2}}{k_{\varphi}^{2}}-\delta=-\frac{\omega_{B}}{\gamma_{r}k_{\varphi}c}.
\end{equation}
This equation describes the wave excitation process by the anomalous
Doppler effect. During the quasi-linear stage of the instability a
diffusion of particles arises not only along but also across the
magnetic field lines. Therefore, plasma particles acquire transverse
momenta (see Eq.12) and as a result the synchrotron mechanism is
switched on.
%
%

The kinetic equation for the distribution function of the resonant
particles can be written in the form (\cite{mach79}, \cite{mach02}):
\begin{eqnarray}
\frac{\partial\textit{f} }{\partial t}+\frac{\partial}{\partial
p_{\parallel}}\left[\left(G_{\parallel}+F_{\parallel}+Q_{\parallel}\right)\textit{f
}\right]+\frac{1}{p_{\parallel}\psi}\frac{\partial}{\partial\psi}\left[\psi\left(G_{\perp}+F_{\perp}\right)\textit{f
}\right]=\nonumber
\\=\frac{1}{\psi}\frac{\partial}{\partial\psi}[\psi\left(D_{\perp,\perp}\frac{\partial}{\partial\psi}+D_{\perp,\parallel}\frac{\partial}{\partial
p_{\parallel}}\right)\textit{f
}\left(p_{\parallel},\psi\right)+\nonumber
 \\
+\frac{\partial}{\partial
p_{\parallel}}\left(D_{\parallel,\perp}\frac{\partial}{\partial\psi}+D_{\parallel,\parallel}\frac{\partial}{\partial
p_{\parallel}}\right)\textit{f }\left(p_{\parallel},\psi\right)],
\end{eqnarray}
where \(G\) is the force responsible for conservation of the
adiabatic invariant \(p_{\perp}^{2}/B(r)=\)const, \(F\)- is the
radiation deceleration force produced by synchrotron emission and
\(Q_{\parallel}\) is the reaction force of the curvature radiation.
They can be written in the form:
\begin{equation}\label{7}
    G_{\perp}=-\frac{mc^{2}}{\rho}\gamma_{r}\psi,\qquad \qquad G_{\parallel}=\frac{mc^{2}}{\rho}\gamma_{r}\psi^{2},
\end{equation}
\begin{equation}\label{8}
    F_{\perp}=-\alpha_{s}\psi\left(1+\gamma_{r}^{2}\psi^{2}\right),\qquad
    F_{\parallel}=-\alpha_{s}\gamma_{r}^{2}\psi^{2},
\end{equation}
\begin{equation}\label{9}
    Q_{\parallel}=-\alpha_{c}\gamma_{r}^{4},
\end{equation}
where $\alpha_{s}=2e^{2}\omega_{B}^{2}/3c^{2}$ and
$\alpha_{c}=2e^{2}/3\rho^{2}$. For the \(\textit{f}_{\parallel}\)
the Eq. (6) will take the form:
\begin{equation}\label{10}
    \frac{\partial\textit{f}_{\parallel} }{\partial t}=\frac{\partial}{\partial
    p_{\parallel}}\left({\left[\alpha_{s}\psi_{0}^{2}\left(\frac{p_{\parallel}}{mc}\right)^{2}+\alpha_{c}\left(\frac{p_{\parallel}}{mc}\right)^{4}-2\pi^{2}\psi_{0}\frac{mc}{p_{\parallel}}r_{e}|E_{k}|^{2}\right]\textit{f}_{\parallel}}\right),
\end{equation}
where \(r_{e}=e^{2}/mc^{2}\), \(\psi_{0}\) is the mean value of the
pitch angle and \(|E_{k}|^{2}\) is the density of electric energy in
the waves.

The resultant spectrum of synchrotron emission is defined by the
shape of distribution function of the resonant particles, which in
turn is the solution of Eq. (10). To define the total flux emitted
by the resonant particles we use the power of a single electron
multiplied by the distribution function of emitting electrons and
integrate the derived expression over the energy distribution (see
\cite{ginz81}).

%
%

%

\section{X-ray spectrum}

We suppose that the measured X-ray spectrum is the result of
 synchrotron emission of primary beam electrons. Let us assume that before the cyclotron instability arises,
 the energy distribution in the beam has a shape:
 \begin{equation}\label{11}
    \textit{f}_{b _{0}}=\frac{n_{b}}{\sqrt{\pi}\gamma_{T}}\textrm{ exp}\left[-\frac{\left(\gamma-\gamma_{b}\right)^{2}}{\gamma_{T}^{2}}\right],
\end{equation}
where \(\gamma_{b}\simeq10^{6}\), \(\gamma_{T}\simeq10\) - is the
half width of the distribution function and \(n_{b}=B/Pce\) is the
density of primary beam electrons, equal to the Goldreich-Julian
density (\cite{gold69}).

The wave excitation leads to a redistribution process of the
particles via the quasi-linear diffusion. Let us compare relative
values of all the three terms in the right-hand side of Eq. (10).
Here we consider that \(\gamma_{b}\psi_{0}\gg1\) (the validity of
this assumption will be estimated after obtaining the expression for
\(\psi_{0}\) ). In this case the equation for the diffusion across
the magnetic field has the form (Malov \& Machabeli 2002):
\begin{eqnarray}
    \frac{\partial\textit{f }_{b_{0}}}{\partial
    t}+\frac{1}{mc\gamma_{r}^{2}\psi}\frac{\partial}{\partial\psi}\left(\psi
    F_{\perp}\textit{f }_{b_{0}}\right)+\frac{1}{mc}\frac{\partial}{\partial\gamma_{b}}\left(F_{\parallel}\textit{f }_{b_{0}}\right)=\nonumber
\\=\frac{1}{m^{2}c^{2}\gamma_{r}^{2}\psi}\frac{\partial}{\partial\psi}\left(\psi
D_{\perp\perp}\frac{\partial\textit{f
}_{b_{0}}}{\partial\psi}\right),
\end{eqnarray}
where
\begin{equation}\label{13}
    D_{\perp\perp}\approx\frac{\pi^{2}e^{2}}{m^{2}c^{3}}\frac{\delta}{\gamma_{b}^{2}}|E_{k}|^{2},
\end{equation}

\begin{equation}\label{14}
    |E_{k}|^{2}=\frac{1}{2}\frac{mc^{2}n_{b}\gamma_{b}c}{\omega}.
\end{equation}

The transversal quasi-linear diffusion increases the pitch-angle,
whereas forces $\overrightarrow{G}$ and $\overrightarrow{F}$ resist
this process, leading to the stationary state
(\(\partial\textit{f}/\partial t=0\)). Then Eq. (12) gives the mean
value of the angle \(\psi\):
\begin{equation}\label{15}
    \psi_{0}=\frac{0.04\gamma_{b}^{3/4}}{BP^{3/4}\gamma_{p}\gamma_{r}^{1/2}}.
\end{equation}
As a result of appearance of the pitch-angles the synchrotron
emission is generated.

Using Eq. (15) we get:
\begin{equation}\label{16}
    \frac{\alpha_{c}\gamma_{r}^{4}}{\alpha_{s}\psi_{0}^{2}\gamma_{r}^{2}}\approx10^{-10}\frac{\gamma_{r}^{3}\gamma_{p}^{2}}{\gamma_{b}^{3/2}P^{1/2}},
\end{equation}
where P is the pulsar rotation period. We suppose that the period of
RXJ1856 is \(P\sim1\)s (this is a typical period for the pulsar with
the age\(\sim10^{6}\)yr). Using the parameter values indicated
above, we will take that the quantity (16) is of order of 10.

Now let us compare the second and the third terms in the right-hand
side of Eq. (10):
\begin{equation}\label{17}
    \frac{2\pi^{2}
    r_{e}\psi_{0}|E_{k}|^{2}/\gamma_{r}}{\alpha_{c}\gamma_{r}^{4}}\approx10^{14}\frac{\gamma_{b}^{11/4}}{B^{2}P^{3/4}\gamma_{p}^{5}\gamma_{r}^{9/2}}.
\end{equation}
We consider that the magnetic field is dipolar. Let us assume that
for the primary-beam electrons the resonance condition is fulfilled
at the distance \(r\simeq10^{9}\)cm then for the magnetic field at
the same distance, we will have \(B\sim10^{2}\)G. For the parameter
values pointed above, it turns out that, the third term of the
Eq.(10) is by six orders of magnitude less than the second one. Then
one gets:
\begin{equation}\label{18}
    \frac{\partial\textit{f}_{\parallel}}{\partial t}=\frac{1}{mc}\frac{\partial}{\partial\gamma_{b}}\left( \alpha_{c}\gamma_{b}^{4}\textit{f}_{\parallel}
    \right).
\end{equation}

For \(\gamma_{b}\psi\ll10^{10}\), a magnetic field inhomogeneity
does not affect the process of wave excitation. The equation which
describes the cyclotron noise level, in this case has the form:
\begin{equation}\label{19}
    \frac{\partial|E_{k}|^{2}}{\partial
    t}=2\Gamma|E_{k}|^{2}\textit{f}_{\parallel},
\end{equation}
where \(\Gamma\) - is the growth rate of instability.
\begin{equation}\label{20}
    \Gamma=\pi\frac{\omega_{p_{res}}^{2}}{\omega_{0}\gamma_{T}}\qquad
    \textrm{if} \qquad\delta\gg\frac{1}{2}\frac{u_{x}^{2}}{c^{2}}
\end{equation}
and
\begin{equation}\label{21}
    \Gamma=\pi\frac{\omega_{p_{res}}^{2}}{2\omega_{0}\gamma_{T}}\frac{u_{x}^{2}}{\delta
    c^{2}}\qquad \textrm{if}
    \qquad\delta\ll\frac{1}{2}\frac{u_{x}^{2}}{c^{2}},
\end{equation}
where the resonant frequency is defined as (\cite{kaz91})
\begin{equation}\label{22}
    \omega_{0}\approx\frac{\omega_{B}}{\delta\gamma_{r}}.
\end{equation}

\begin{figure}
   \centering
\includegraphics[width=9cm]{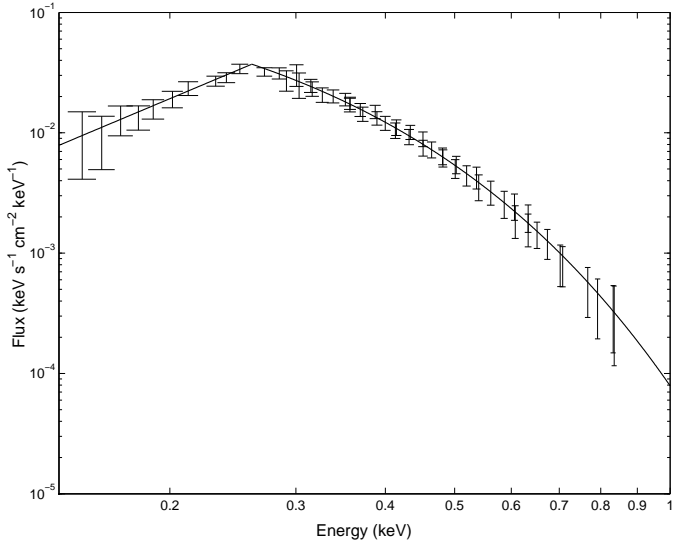}
      \caption{The \emph{Chandra} LETG X-ray spectrum of RX J1856.5-3754
      fitted with the model.
              }
         \label{Figmodel}
   \end{figure}

In this case we can use Eq. (21) for the growth rate and after
substituting Eq. (19) in (18), we will take:
\begin{equation}\label{23}
    \frac{\partial}{\partial t}\left(\textit{f}_{\parallel}-\frac{\alpha_{c}}{m}\frac{\omega_{B}\gamma_{T}c}{\omega_{p_{res}}^{2}u^{2}}\frac{\partial}{\partial\gamma_{b } }
    \gamma_{b}^{3}\ln|E_{k}|^{2}\right)=0.
\end{equation}
From this equation it is easy to find the distribution function:
\begin{equation}\label{24}
    \textit{f}_{\parallel_{b}}=\textit{f}_{\parallel_{b}0}+C\gamma_{b}^{2}.
\end{equation}
We suppose that the right slope of the measured X-ray spectrum is
the result of synchrotron emission of the primary-beam electrons,
with the distribution function (24), see Fig.~\ref{Figmodel}.

In \cite{malov02} it has been shown that it is possible to form the
stationary distribution. If we assume \(\partial
\textit{f}_{\parallel}/\partial t=0\), then from Eq. (18) one finds:
\begin{equation}\label{25}
    \textit{f}_{\parallel _{b}}\propto\gamma_{b}^{-4}.
\end{equation}
We consider that after the quasi-linear evolution stage of the
instability, by achieving the stationary state, the radiation
density is significant and the self-absorption effects begin to play
the main role. Synchrotron self-absorption, redistributes the
emission spectrum and in the domain of relatively low frequencies,
we have (\cite{zhel77}):
\begin{equation}\label{26}
    I(\nu)\propto\nu^{5/2}.
\end{equation}
This emission corresponds to energy interval (0.15-0.26 keV) of the
measured X-ray spectrum (see Fig.~\ref{Figmodel}).

The frequency of the maximum of measured X-ray spectrum corresponds
to the intersection point of theoretical curves, fitted with the
observed data (\(\nu\simeq6\cdot10^{16}\)Hz). On the other hand, the
frequency of the maximum of synchrotron spectrum of a single
electron is (\cite{pacho70}):
\begin{equation}\label{27}
    \nu_{m}\simeq0.07\frac{eB}{mc}sin\psi\gamma_{b}^{2}.
\end{equation}

Using Eq. (15) for \(\psi\ll1\), we find that the frequency of
maximum of the spectrum turns out to be not dependent on magnetic
field. Consequently, we can estimate from Eq. (27) the rotation
period of RXJ1856. Estimations show that \(P\sim1\)s, i.e. the
primitive conjecture proves to be true. The same can be found about
two other assumptions done at the beginning of our computations,
since \(\psi_{0b}\sim10^{-3}\ll1\) and
\(\gamma_{b}\psi_{0b}\sim10^{3}\).


\section{Optical spectrum}

Now let us consider the observed optical emission of RXJ1856. We
suppose that this emission is related with electrons of a tail, with
an average Lorentz-factor \(\gamma_{t}\simeq10^{4}\). Let us assume
that the initial distribution of particles of the tail has the
following form (see Fig.~\ref{Fig2}):
\begin{equation}\label{28}
    \textit{f}_{t _{0}}\propto\gamma^{-4}.
\end{equation}

As in the previous case, one needs to estimate the contribution of
different terms in the right-hand side of Eq. (10). We suppose that
the condition \(\gamma_{t}\psi_{0}\gg1\) is fulfilled and we can use
Eq. (15) for the mean value of angle \(\psi\) in this case too. For
values of parameters indicated above we take that, the first term
exceeds the second one by five orders of magnitude. Now let us
compare the first and the third terms:
\begin{equation}\label{29}
    \frac{2\pi^{2}
    r_{e}\psi_{0}|E_{k}|^{2}/\gamma_{t}}{\alpha_{s}\psi_{0}^{2}\gamma_{r}^{2}}\approx1.3\cdot10^{4}\frac{\gamma_{b}^{5/4}}{B^{2}P^{5/4}\gamma_{p}^{3}\gamma_{t}^{3/2}}.
\end{equation}
We suppose that for the electrons of the tail, the cyclotron
resonance condition is fulfilled at the distance
\(r\simeq3\cdot10^{9}\)cm and for the value of the magnetic field we
take \(B\sim1\)G. Finally, after substituting values in Eq. (29) we
take that the third term is less than the first one by an order of
magnitude.

Consequently, instead of Eq. (10), we will have:
\begin{equation}\label{30}
    \frac{\partial\textit{f}_{\parallel}}{\partial
    t}+\frac{2\pi^{2}\psi_{0}r_{e}}{mc}\frac{\partial}{\partial\gamma}\left(\frac{|E_{k}|^{2}}{\gamma}\textit{f}_{\parallel}
    \right)=0.
\end{equation}
In this case we can use Eq. (20) for the growth rate of
instability and from Eqs. (19) and (30) one gets:
\begin{equation}\label{31}
    \frac{\partial}{\partial
    t}\left(\textit{f}_{\parallel}+\frac{\pi^{2}e^{2}\omega_{B}\gamma_{T}}{m^{2}c^{3}\delta\omega_{p_{t}}^{2}}\psi_{0}\frac{\partial}{\partial\gamma}\frac{|E_{k}|^{2}}{\gamma^{2}}\right)=0.
\end{equation}
From this equation we can find the expression for $|E_{k}|^{2}$ at
the final moment of the quasi-linear stage. If we take into account
that the density of the particles behaves as $1/r^3$ ($r$ is the
distance from the pulsar), one can neglect \(
\textit{f}_{\parallel_{t}}\) in comparison with
\(\textit{f}_{\parallel_{t}0}\) and from Eq. (31) one finds:
\begin{equation}\label{32}
   |E_{k}|^{2}\approx\frac{m^{2}c^{3}\delta\omega_{p_{t}}^{2}}{2\pi^{2}e^{2}\omega_{B}\gamma_{T}\psi_{0}\gamma}.
\end{equation}

After achieving the stationary mode, the distribution function of
particles of the tail can be found from Eq. (30) if one sets
\(\partial\textit{f}_{\parallel}/\partial t=0\), then we will take:
\begin{equation}\label{33}
    \textit{f}_{\parallel}\propto|E_{k}|^{-2}\gamma.
\end{equation}
Using Eq. (32) for the \(|E_{k}|^{-2}\) we find:
\begin{equation}\label{34}
    \textit{f}_{\parallel_{t}}\propto\gamma^{2}.
\end{equation}
The resultant theoretical spectrum is well matched with the measured
one. For parameters used above, we find that indeed
\(\gamma_{t}\psi_{0t}\sim10^{3}\gg1\)
(\(\psi_{0t}\simeq3\cdot10^{-1}\)).

The frequency of the maximum in synchrotron emission of a single
electron is defined by Eq. (27). For the parameter values used
above, we will take \(\nu_{m}(Optic)\simeq2\cdot10^{14}\)Hz. This
quantity comes in the domain that corresponds the observed optical
data.

%

\section{The effectiveness of the cyclotron mechanism}

Now let us estimate the frequencies of original waves, excited
during the cyclotron resonance. Using the same parameters, one can
see from Eq. (22) that the frequencies corresponding to the beam and
tail components are \(\nu\approx3\cdot10^{9}\)Hz and
\(\nu\approx5\cdot10^{8}\)Hz respectively, i.e. during the cyclotron
resonance in both cases the radio waves are excited. Though this
waves are generated earlier than X-ray and optical emission and
therefore the radio emission might pass by line of sight. In this
case the radio emission will not get to observer. That is one of
possible explanations why the radio emission is not detected from
this object.

For effective generation of waves it is essential that time, during
which the particles give energy to waves should be more than
$1/\Gamma$. Generated radio waves propagate practically in straight
lines, whereas the field line of the dipole magnetic field deviates
from its initial direction and the angle
$\theta=k_{\parallel}/k_{\varphi}$ grows (\(\theta\) is the angle
between the wave line and the line of dipole magnetic field). On the
other hand, the resonance condition (5) imposes limitations on
\(\theta\approx\textrm{max} \{\sqrt{\delta},\frac{u_{x}}{c}\}\)
(\cite{kaz91}), i.e. particles can resonate with the waves
propagating in a limited range of angles. Obviously the  Eq. (5)
will be fulfilled until then (\cite{lyu99}):
\begin{equation}\label{35}
    \frac{c}{\Gamma}\lesssim\theta\rho,
\end{equation}
where $c\Gamma^{-1}$ is the growth length and $\theta\rho$ is the
length of the wave-particle interaction. For the beam particles from
Eq. (35) it follows that \(\rho\gtrsim3\cdot10^{9}\)cm. As the
cyclotron instability arises at distances $r\simeq10^{9}$cm for the
beam electrons, this result means that time of wave interaction with
the resonant particles is quite enough for particles to acquire the
pith-angles $\psi_{0}$ (Eq. 15), which automatically leads to the
generation of synchrotron emission.

The total energy available for the conversion into pulsar emission
is of the order of the energy of primary beam particles flowing
along the open filed lines of the pulsar magnetosphere:
\begin{equation}\label{36}
    \dot{E} \simeq n_{b_{0}} \pi R_{pc}^{2}\gamma_{b}mc^{3},
\end{equation}
where $n_{b_{0}}$ is the Goldreich-Julian density at the pulsar
surface and $R_{pc}$ is the radius of the polar cap. The estimations
show $\dot{E}\simeq5\cdot10^{32}$erg/s, which is quite enough to
explain the observed luminosity of RXJ1856.

%
%

%

%

\section{Conclusions}

\begin{enumerate}
\item In the present paper we assume that the emission of RXJ1856 is
generated by the synchrotron mechanism. The main reason of wave
generation in outer parts of the pulsar magnetosphere is the
cyclotron instability. During the quasi-linear stage of the
instability a diffusion of particles arises as along, also across
the magnetic field. Plasma particles acquire pitch angles and begin
to rotate along the Larmor orbits. Synchrotron emission is the
result of the appearance of pitch angles.

\item The measured X-ray spectrum is the result of synchrotron emission of
primary-beam electrons, generated at distances \(r\simeq10^{9}\)cm,
when the optical emission is related with secondary plasma
particles, particularly the tail electrons. The optical spectrum is
produced by synchrotron emission of the tail electrons, generated at
distances \(r\simeq3\cdot10^{9}\)cm. The predictable characteristic
frequencies \(\nu_{m}(X-ray)\simeq6\cdot10^{16}\)Hz and
\(\nu_{m}(Optic)\simeq2\cdot10^{14}\)Hz come in the same domains as
the measured spectra.

\item The observed optical emission reveals an intensity six times larger than
the X-ray one. The X-ray emission is generated earlier in contrast
to optical emission. Thus, an observer may receive only part of the
radiation emitted in the X-ray domain. We suggest that this must be
the reason of lower intensity of the X-ray spectrum.

\item In view of the fact that the aligned
rotator model is used, the radio emission covers a large distance in
the pulsar magnetosphere. So there is a high probability for it to
come in the cyclotron damping range
\(\omega-k_{\varphi}V_{\varphi}-k_{x}u_{x}-\frac{\omega_{B}}{\gamma_{r}}=0\).
In this case the radio emission will not reach an observer. Another
explanation of the lack of radio emission from this object is that
the X-ray and optical emission, resulted from the quasi-linear
diffusion, occur in a region further away from that of the radio
emission and it miss our line of sight. Nevertheless the detection
of radio emission from RXJ1856 would be strong argument to our
model.

\item We give estimations of effectiveness of the cyclotron
mechanism, which in this case means the fulfillment of the following
condition \(\rho\gtrsim3\cdot10^{9}\)cm. As the instability develops
at distances $r\simeq10^{9}$cm, then follows that the excited waves
lie in the resonant region sufficiently long time, which is quite
enough for particles to acquire perpendicular momentum and generate
radiation.

\end{enumerate}

\begin{acknowledgements}
We are grateful to Zaza Osmanov for helpful advises and Xiaoling
Zhang for providing the X-ray data. This research was supported by
Georgian NSF grant ST06/4-096.
\end{acknowledgements}


\begin{thebibliography}{}

  \bibitem [Arons 1981]{arons81}Arons, J. 1981, In: Proc. Varenna
Summer School and Workshop on Plasma Astrophysics, ESA, p.273
\bibitem [Braje \& Romani 2002] {braje02} Braje, T. M., Romani,
R. W., 2002, ApJ, 580, 1043
\bibitem [Burwitz et al. (2001)] {burw01} Burwitz, V., Zavlin, V.
E., Neuh\"{a}user, R., Predehl, P., Tr\"{u}mper, J. Brinkman, A. C.,
2001, A\&A, 379, L35
\bibitem [Burwitz et al. (2003)] {burw03} Burwitz, V., Haberl, F.
Neuh\"{a}user, R., Predehl, P., Tr\"{u}mper, J. Zavlin, V. E., 2003,
A\&A, 399, 1109
\bibitem [Ginzburg 1981] {ginz81} Ginzburg, V. L., "Teoreticheskaia Fizika i Astrofizika" ,Nauka, Moskva 1981
\bibitem [Goldreich \& Julian 1969] {gold69} Goldreich, P.,
Julian, W. H., 1969, ApJ, 157, 869
\bibitem[Ho et al. 2007] {ho07} Ho, W. C. G., Kaplan, D. L., Chang,
P., Adelsberg, M., Potekhin, A. Y., 2007, MNRAS 375, 281H
\bibitem[Kazbegi, Machabeli \& Melikidze 1991] {kaz91} Kazbegi, A. Z., Machabeli,
G. Z., Melikidze, G. I., 1991, MNRAS 253,377
\bibitem [Lai \& Salpeter 1997] {lai97} Lai, D., Salpeter, E. E.,
1997, ApJ, 491, 270
\bibitem [Lai 2001] {lai01} Lai, D., 2001, Rev. of Mod. Phys.
73,629
\bibitem [Lominadze, Machabeli \& Mikhailovskii (1979)] {lom79} Lominadze, D. G.,
Machabeli, G. Z., Mikhailovskii, A. B., 1979, Fiz. Plaz., 5,1337
\bibitem [Lyutikov, Blandford \& Machabeli 1999] {lyu99} Lyutikov,
M., Blandford, R. D., Machabeli, G. Z., 1999, MNRAS 305, 338L
\bibitem [Machabeli \& Usov 1979] {mach79} Machabeli, G. Z., Usov,
V. V. 1979 AZh Pis'ma, 5, 445
\bibitem [Machabeli et al. 2002] {mach02} Machabeli, G. Z., Luo,
Q., Vladimirov, S, V., Melrose, D. B., 2002, Phys. Rev., E 65,
036408
\bibitem [Malov \& Machabeli (2002)] {malov02} Malov, I, F.,
Machabeli, G. Z., 2002, Astronomy Reports, Vol. 46, Issue 8, p.684
\bibitem [Pacholczyk et al. 1970] {pacho70} Pacholczyk, A. G.
1970, Radio Astrophysics (San Francisco: W. H. Freeman)
\bibitem [Pavlov et al. (1996)] {pavlov96} Pavlov, G. G., Zavlin, V.
E., Tr\"{u}mper, J., Neuh\"{a}user, R., 1996, ApJ, 472, L33
\bibitem [Pavlov 2000] {pavlov00} Pavlov, G. G., 2000, Talk at the
ITP/UCSB workshop "Spin and Magnetism of Young Neutron Stars"
\bibitem [Pavlov, Zavlin \& Sanwal 2002] {pavlov02} Pavlov, G, G., Zavlin, V.
E., Sanwal, D., in Neutron Stars and Supernova Remnants. Eds. W.
Becher, H. Lesch, \& J. Tr\"{u}mper, 2002, MPE Report 278,273
\bibitem [Pons et al. (2002)] {pons02} Pons, J. A., Walter, F. M.,
Lattimer, J. M., Prakash, M., Neuha\"{a}user, R., An, P., 2002, ApJ,
564, 981
\bibitem [Rajagopal et al. 1997] {raja97} Rajagopal, M., Romani,
R. W., Miller, M. C., 1997, ApJ, 479, 347
\bibitem [Turolla, Zane \& Drake (2004)] {turolla04} Turolla, R., Zane, S. Drake, J. J., 2004, ApJ, 603, 265
\bibitem [Walter et al. 1996] {walt96} Walter, F. M., Wolk, S. J.,
Neuh\"{a}user, R., 1996, Nature 379,233
\bibitem [Zavlin \& Pavlov 2002] {zav02} Zavlin, V. E., Pavlov, G.
G., 2002 in Neutron Stars and Supernova Remnants, Eds. Becker, W.,
Lesch, H., Tr\"{u}mper, J., MPE Report 278, p.261
\bibitem [Zhelezniakov 1977] {zhel77} Zhelezniakov, V. V., "Volni v Kosmicheskoi Plazme" Nauka, Moskva 1977

\end{thebibliography}
\end{document}